\documentclass[aps,twocolumn,preprintnumbers,nofootinbib,superscriptaddress]{revtex4}
\usepackage{amsmath, amssymb} \usepackage{graphicx} \usepackage{amsfonts}
\usepackage{array} \usepackage{amsthm} \usepackage{bm}
\usepackage{palatino} \usepackage{mathpazo} %\usepackage{mathptmx}
\usepackage{supertabular}

\usepackage[breaklinks]{hyperref}
\usepackage{color}

\newcommand{\nn}{\nonumber}
\newcommand{\be}{\begin{equation}}
\newcommand{\ee}{\end{equation}}
\newcommand{\ba}{\begin{eqnarray}}
\newcommand{\ea}{\end{eqnarray}}
\newcommand{\bal}{\begin{align}}
\newcommand{\eal}{\end{align}}

\newcommand{\e}{{\rm e}}
\newcommand{\dd}{{\rm d}}

\newcommand{\bb}{\bibitem}

\newcommand{\om}{\omega}
\newcommand{\al}{\alpha}

\newcommand{\ga}{\gamma}

\newcommand{\ep}{\epsilon}
\newcommand{\si}{\sigma}
\newcommand{\ta}{\theta}

\newcommand{\Om}{\Omega}

\newcommand{\bw}{\begin{widetext}}
\newcommand{\ew}{\end{widetext}}

\begin{document}
%\begin{flushleft}???
%\end{flushleft}
\title{Accretion of rotating fluids onto stationary solutions}

\author{Mustapha Azreg-A\"{\i}nou}%\email{?????}
\affiliation{Ba\c{s}kent University, Engineering Faculty, Ba\u{g}l\i ca Campus, Ankara, Turkey}

%\date{}

\begin{abstract}
We consider a general stationary solution and derive the general laws for accretion of rotating perfect fluids.
For non-degenerate and degenerate Fermi and Bose fluids we derive new effects that mimic the center-of-mass-energy effect of two colliding particle in the vicinity of horizons. Non-degenerate fluids see their chemical potential grow arbitrarily and ultra-relativistic Fermi fluids see their specific enthalpy and Fermi momentum grow arbitrarily too while the latter vanishes gradually for non-relativistic Fermi fluids. For degenerate Bose fluids two scenarios remain possible as the fluid approaches a horizon: a) The Bose-Einstein condensation ceases or b) the temperature drops gradually down to zero. The critical flow is also investigated.
\end{abstract}

%\pacs{}

\maketitle

\section{What is accretion?\label{secw}}
Given a metric solution of spacetime, a geodesic motion is the process by which a massive or massless test particle ``falls" freely. The fall motion may be bounded (circular motion or else) or unbounded (scattering motion). The free fall of the test particle does not disturb, affect, or modify the given geometry: No back reaction effects are taken into consideration. Moreover, no group motion is treated in geodesic motion. Back reaction effects are present in the calculation of gravitational self forces where the motion of the ``small" body is still seen as geodesic in the perturbed metric.

Accretion is an advanced state of motion. It describes group motion with and without back reaction and it is generally non-geodesic. By group motion it is meant that the accreting matter is modeled by a fluid and that each fluid element encompasses a) a sufficiently large number of particles to be described statistically by an average pressure, average temperature, average particle number density, and average energy density; b) a sufficiently small number of particles compared to the whole system (accreting matter, atmosphere, etc). These conditions are easily met in astrophysics and atmospheric motion. The presence of a gradient of pressure, which is a sort of fluid group self force, renders the accretion motion non-geodesic.

In accretion motion back reaction effects are taken into consideration in numerical and simulation analyses~\cite{Dur,Matteo,Russell}. Full analytical treatments~\cite{acc00,acc0,acc1,acc2,Bertozzi} drop back reaction effects for simplicity and cognitive treatments~\cite{Shapiro,basic,Chat,string,aafj,aafs,add} may include emission effects and neglect back reaction too.

All the above-mentioned treatments make common simplificative physical assumptions of symmetry concerning both the given background geometry and the fluid. They assume the background metric to remain stationary and time-independent during accretion~\cite{houches}. The analysis remains valid for accretion time, larger than free-fall time, and much smaller than the ratio mass/(mass rate change) of the star.

In this work we will keep using the standard set of simplificative assumptions to describe the accretion of rotating perfect fluids onto rotating black holes with no back reaction or emission effects. In Sec.~\ref{secm} we present the accretion model and in sec.~\ref{secg} we derive the general equations for accretion of a perfect fluid onto a stationary rotating black hole. We keep using a general stationary metric throughout the paper. In Sec.~\ref{seceb} we investigate the end-behavior of accretion as the fluid approaches horizons and other end-points. Sec.~\ref{secndd} is devoted to important applications concerning the accretion of non-degenerate as well as degenerate Fermi and Bose fluids. In Sec.~\ref{seccf} we derive the conditions of a critical flow and the corresponding critical points. An appendix section has been added to complete the discussion of, and to derive some equations pertaining to, Sec.~\ref{seceb}. We conclude in Sec.~\ref{secc}.

%Analytical accretion~\cite{acc00}-\cite{Bertozzi}

%Cognitive~\cite{Shapiro}-\cite{add}

\section{Model for analytical accretion\label{secm}}
Consider a stationary rotating generic metric of the form
\begin{equation}\label{0m1}
\hspace{-1.4mm}\dd s^2=g_{tt}\dd t^2+2g_{t\phi}\dd t\dd \phi+g_{\phi\phi}\dd \phi^2+g_{rr}\dd r^2+g_{\ta\ta}\dd \ta^2,
\end{equation}
admitting two Killing vectors, one is timelike $\xi_t^{\mu}=(1,\,0,\,0,\,0)$ and the other is spacelike $\xi_{\phi}^{\mu}=(0,\,0,\,0,\,1)$. The metric components $g_{\mu\nu}$ are functions of the radial and polar coordinates ($r,\,\ta$) and $g_{t\phi}$ vanishes identically if rotation is suppressed. We do not assume any specific asymptotic end-behavior (asymptotic flatness, de Sitter, or anti-de Sitter behavior), for the analysis we intend to give in this work applies whatever the end-behavior of the black hole is.

The metric may have an event horizon, a cosmological horizon, and possibly other horizons all denoted by $r_h$. In any case we will be concerned with the regions of the three-space where the Killing vector $\xi_t^{\mu}$ is timelike. These are the regions accessible to the fluid flow and to observers.

We choose the signature ($-,\,+,\,+,\,+$) for~\eqref{0m1}. In all our mathematical expressions we will keep using absolute values so that the formulas remain valid if the signature ($+,\,-,\,-,\,-$) is adopted.

\subsection{Zero-angular-momentum observers}
Setting
\begin{equation}\label{0m2}
\hspace{-1.7mm}\om (r,\ta)\equiv -g_{t\phi}/g_{\phi\phi}>0, \; D(r,\ta)\equiv g_{t\phi}^2-g_{tt}g_{\phi\phi}\geq 0,
\end{equation}
we bring the metric~\eqref{0m1} to the following useful form
\begin{equation}\label{0m3}
\hspace{-2.2mm}\dd s^2=\frac{D}{-g_{\phi\phi}}\,\dd t^2+g_{\phi\phi}(\dd \phi-\om\dd t)^2+g_{rr}\dd r^2+g_{\ta\ta}\dd \ta^2.
\end{equation}
One can always make it such that $\om >0$ (for instance, by changing the positive direction of the $z$ axis, which coincides with the axis of symmetry). However, the inequalities in~\eqref{0m2} are supposed to hold in the regions of the three-space where the Killing vector $\xi_t^{\mu}$ is timelike.

It is straightforward to show that $\om$ is the angular velocity of the zero-angular-momentum observers (ZAMOs)~\cite{w2}. We choose a reference frame $(e_t,\,e_r,\,e_{\ta},\,e_{\phi})$ dual to the 1-forms defined in~\eqref{0m3}: $\om^t\equiv -\sqrt{D/|g_{\phi\phi}|}\,\dd t$, $\om^r\equiv \sqrt{|g_{rr}|}\,\dd r$, $\om^{\ta}\equiv \sqrt{|g_{\ta\ta}|}\,\dd \ta$, $\om^{\phi}\equiv \sqrt{|g_{\phi\phi}|} (\dd \phi-\om\dd t)$. The frame $(e_t,\,e_r,\,e_{\ta},\,e_{\phi})$ is given by
\begin{align}
&e^{\mu}_t=\sqrt{\frac{|g_{\phi\phi}|}{D}}\,(1,\,0,\,0,\,\om),
& & \hspace{-1.5mm} e^{\mu}_r=\sqrt{\frac{1}{|g_{rr}|}}\,(0,\,1,\,0,\,0),\nn\\
\label{m4}&e^{\mu}_{\ta}=\sqrt{\frac{1}{|g_{\ta\ta}|}}\,(0,\,0,\,1,\,0),
& & \hspace{-1.5mm} e^{\mu}_{\phi}=\sqrt{\frac{1}{|g_{\phi\phi}|}}\,(0,\,0,\,0,\,1).
\end{align}

Consider a ZAMO with four-velocity $\bar{u}^{\mu}=e^{\mu}_t$~\eqref{m4}. Relative to the $\bar{u}^{\mu}$ frame, the four-velocity vector of the fluid,
\begin{equation}\label{m5}
u^{\mu}=\Big(\frac{\dd t}{\dd \tau},\,\frac{\dd r}{\dd \tau},\,\frac{\dd \ta}{\dd \tau},\,\frac{\dd \phi}{\dd \tau}\Big)=(u^t,u^r,u^{\ta},u^{\phi}),
\end{equation}
with $\tau$ being the proper time, expands as
\begin{equation}\label{m6}
    u^{\mu}=\ga (e^{\mu}_t+ve^{\mu}_r+\hat{u}e^{\mu}_{\ta}+ue^{\mu}_{\phi}),
\end{equation}
where we have set
\begin{equation}\label{m7}
    V^2\equiv v^2+\hat{u}^2+u^2\;\text{ and }\;\ga\equiv \frac{1}{\sqrt{1-V^2}}.
\end{equation}
On comparing~\eqref{m5} and~\eqref{m6} and using~\eqref{m4} it is easy to establish that ($v,\,\hat{u},\,u$) are the three-dimensional linear components of the three-velocity of the fluid relative to the $\bar{u}^{\mu}$ reference frame and $V$ is the (relative) three-speed. The components ($v,\,\hat{u},\,u$) along with $u^t$ are given by
\begin{multline}\label{0m7}
\hspace{-3mm}v=\sqrt{\frac{g_{rr}g_{\phi\phi}}{D}}~\frac{\dd r}{\dd t},\;\hat{u}=\sqrt{\frac{g_{\ta\ta}g_{\phi\phi}}{D}}~\frac{\dd \ta}{\dd t},\;u=\frac{|g_{\phi\phi}|}{\sqrt{D}}(\Om-\om), \\u^t=\gamma \sqrt{\frac{|g_{\phi\phi}|}{D}},\qquad
\end{multline}
where $\Om\equiv\dd \phi/\dd t$ is the differential angular velocity of the fluid.

\subsection{The stress-energy tensor}
From now on, we consider an accreting perfect fluid containing a single particle species.

In arbitrary coordinates, we consider a general observer (not necessarily co-moving with the fluid) with four-velocity $U^{\mu}$. We reserve the notation with large characters to physical and geometrical entities measured by, or pertaining to, this observer. Such a general observer can decompose the stress-energy tensor (SET) of the fluid in the form
\begin{equation}\label{of3}
T_{\mu\nu}=(E+P)U_{\mu}U_{\nu}+Pg_{\mu\nu}+\Pi_{\mu\nu},
\end{equation}
where $E=U^{\mu}U^{\nu}T_{\mu\nu}$ is the relativistic energy density as measured by the observer with four-velocity $U^{\mu}$, $\Pi_{\mu\nu}$ is the traceless anisotropic pressure tensor satisfying $U^{\mu}U^{\nu}\Pi_{\mu\nu}=0$, and $P=-(g^{\mu\nu}-U^{\mu}U^{\nu})T_{\mu\nu}/3$ is the average relativistic isotropic pressure as measured by this observer. In the rest frame of the perfect fluid we assume the latter to be isotropic; that is, the observer co-moving with the fluid ($U^{\mu}=u^{\mu}$) measures a vanishing anisotropic pressure tensor $\pi_{\mu\nu}\equiv 0$ and decomposes the SET of the fluid as
\begin{equation}\label{of3a}
T_{\mu\nu}=(\ep+p)u_{\mu}u_{\nu}+pg_{\mu\nu},
\end{equation}
where $\ep$ and $p$ are the energy density and pressure in the rest frame of the perfect fluid.

The assumption of being isotropic holds in the $u^{\mu}$ frame only. It is easy to show that in the $\bar{u}^{\mu}$ frame the transverse pressure of the fluid is not isotropic and splits into two components $p_{\ta}\neq p_{\phi}$ defined as $e^{\mu}_{\ta}e^{\nu}_{\ta}T_{\mu\nu}$ and $e^{\mu}_{\phi}e^{\nu}_{\phi}T_{\mu\nu}$, respectively. In the latter frame, the SET splits as
\begin{equation*}
    T_{\mu\nu}=(\bar{\ep}+\bar{p})\bar{u}_{\mu}\bar{u}_{\nu}+\bar{p}g_{\mu\nu}+\bar{\pi}_{\mu\nu},
\end{equation*}
which results in
\begin{equation*}
   \bar{\ep}=\ga^2 (\ep +V^2p),\qquad \bar{p}=p+\frac{V^2\ga^2(\ep+p)}{3},
\end{equation*}
and, for instance,
\begin{align*}
&e^{\mu}_{r}e^{\nu}_{r}\bar{\pi}_{\mu\nu}=\Big(v^2-\frac{V^2}{3}\Big)\ga^2(\ep+p),\\
&e^{\mu}_{\ta}e^{\nu}_{\ta}\bar{\pi}_{\mu\nu}=\Big(\hat{u}^2-\frac{V^2}{3}\Big)\ga^2(\ep+p),\\
&e^{\mu}_{\phi}e^{\nu}_{\phi}\bar{\pi}_{\mu\nu}=\Big(u^2-\frac{V^2}{3}\Big)\ga^2(\ep+p).
\end{align*}
It is clear from these relations and $\bar{u}^{\mu}\bar{u}^{\nu}\bar{\pi}_{\mu\nu}=0$ that $\bar{\pi}^{\mu}{}_{\mu}=0$.

\subsection{Assumptions and choice of a ZAMO\label{assump}}
It is usually assumed that the accreting matter moves within a narrow planar disk that one chooses to be the $\ta=\pi/2$ plane~\cite{Frank}. The rotation concentrates the streamlines in the equatorial plane~\cite{MHD} so that the thickness of the disk is assumed to be much smaller than its extent in the $\ta=\pi/2$ plane, in that, we neglect any variation with respect to $\ta$:
\begin{equation}\label{ta}
\partial_{\ta}F|_{\ta=\pi/2}=0,
\end{equation}
where $F$ represents some physical or geometrical entity. The disk is supposed to surround the black hoke in such a way that one generally drops any dependence on the azimuthal angle $\phi$.

To perform analytical treatment of accretion, it is also commonly assumed that the fluid is in a steady state for most of the accretion time. Thus, the four-velocity vector and the thermodynamic properties of the fluid and related physical entities are considered to be independent of time.

These last two paragraphs ensure that the fluid motion is endowed with the same symmetry properties of the spacetime.

If the metric~\eqref{0m1} describes a black hole, it has a number of horizons all denoted by $r_h$ where (in the $\ta=\pi/2$ plane, $r_h$ is a constant)
\begin{equation}\label{ta2}
1/g_{rr}(r_h,\ta)=0\quad\text{ and }\quad 0< |g(r_h,\ta)|<\infty,
\end{equation}
that is, the metric is regular at $r=r_h$. Since the absolute value of the determinant $|g|$ is given by $|g|=g_{rr}g_{\ta\ta}D$~\eqref{0m3}, $D$ must go to zero as $1/g_{rr}$ in the limit $r\to r_h$ to ensure that $|g|$ remains regular there
\begin{equation}\label{ta3}
    D(r_h,\ta)\equiv D_h=0,
\end{equation}
where we have introduced the notation $F(r_h)\equiv F_h$ with $F$ being some physical or geometrical entity.

The metric~\eqref{0m1} may admit a static limit, which is the 2-surface on which the timelike Killing vector $\xi_t^{\mu}$ becomes null, corresponding to
\begin{equation}\label{ta4}
g_{tt}(r_{\text{static}},\ta)=0.
\end{equation}
Observers can remain static only for $g_{tt}<0.$\footnote{This is the case if the signature of the metric~\eqref{0m1} is ($-,\,+,\,+,\,+$); for a ($+,\,-,\,-,\,-$) signature the statement reads $g_{tt}>0$.} Whether such a static limit exists or not, a convenient reference frame for describing the fluid flow in the regions of the three-space where the Killing vector $\xi_t^{\mu}$ is timelike is a ZAMO $\mathcal{O}_\circ$ with four-velocity $\bar{u}^{\mu}=e^{\mu}_t$~\eqref{m4} located at a constant radius $r=\mathrm{const}$ and a constant angle $\theta=\pi/2$, as is always the case in accretion problem~\cite[p. 114]{houches}. Relative to this stationary observer $\mathcal{O}_\circ$, $\hat{u}\equiv 0$ and ($v,\;u$) are given by~\eqref{0m7} where the rhs's are evaluated at $\ta=\pi/2$.

\section{General equations for accretion of rotating fluids\label{secg}}
We define the current density by $J^{\mu}=n u^{\mu}$ where $n$ is the particle density, precisely the baryon number density in the fluid rest frame (recall that we consider an accreting perfect fluid containing a single particle species). From the  particle conservation law, we have that the divergence of the current density is conserved
\begin{equation}\label{g1}
\nabla_{\mu}(n u^{\mu})=\frac{1}{\sqrt{|g|}}\partial_{\mu}\big(\sqrt{|g|}n u^{\mu}\big)=0,
\end{equation}
where $\nabla_{\mu}$ is the covariant derivative. In the $\ta=\pi/2$ plane, this reduces using~\eqref{ta} to $\partial_{r}(\sqrt{|g|}n u^r)=0$ or, upon integrating, to
\begin{equation}\label{g1b}
\big(\sqrt{g_{rr}g_{\ta\ta}D}\big)\big|_{\ta=\pi/2}n u^{r}=C_{1},
\end{equation}
where we have used $|g|=g_{rr}g_{\ta\ta}D$. It is understood that ($n,\,u^r$) are evaluated at $\ta=\pi/2$. Here $C_{1}$ is a numerical constant of integration.

Reversing~\eqref{0m7} (or, equivalently, using the normalization condition $u^{\mu}u_{\mu}=1$), we obtain
\begin{equation}\label{g2}
(u^t)^2=\frac{|g_{\phi\phi}|}{D(1-V^2)},\quad (u^r)^2=\frac{v^2}{|g_{rr}|(1-V^2)}.
\end{equation}

A theorem of relativistic hydrodynamics~\cite{Rezzolla} states that the scalar $hu_{\mu}\xi^{\mu}$ is conserved along the trajectories of the fluid,
\begin{equation}\label{t3}
u^{\nu}\nabla_{\nu}(hu_{\mu}\xi^{\mu})=0,
\end{equation}
where $\xi^{\mu}$ is a Killing vector of spacetime (in the case of the metric~\eqref{0m1}, $\xi^{\mu}=\xi_t^{\mu}$ or $\xi^{\mu}=\xi_{\phi}^{\mu}$) and $h$ is the specific enthalpy (enthalpy per particle) defined by
\begin{equation}\label{ent}
h=\frac{\epsilon+p}{n}.
\end{equation}
This theorem stems from the fact that the fluid motion is endowed with the same symmetry properties of the spacetime (see assumptions made in the first two paragraphs of Sec.~\ref{assump}). Consider the timelike Killing vector $\xi_t^{\mu}$ of the metric~\eqref{0m1}, Eq.~\eqref{t3} yields $\partial_{r}(hu_{t})=0$ (where we have used~\eqref{ta} and $u^{\ta}=0$ on the $\ta=\pi/2$ plane). With $u_t=g_{tt}u^t+g_{t\phi}u^{\phi}=(g_{tt}+\Om g_{t\phi})u^t$, this finally yields
\begin{equation}\label{t4}
h(g_{tt}+\Om g_{t\phi})u^t=C_2,
\end{equation}
where $C_2$ is a numerical constant of integration. On considering the spacelike Killing vector $\xi_{\phi}^{\mu}$, we obtain
\begin{equation}\label{t5}
h(\Om -\om)g_{\phi\phi}u^t=C_3,
\end{equation}
where $C_3$ is a numerical constant of integration. In Eqs.~\eqref{t4} and~\eqref{t5} it is understood that all the physical and geometrical entities are evaluated at $\ta=\pi/2$.

The meaning of the constants ($C_2,\,C_3$) is as follows. If accretion were a geodesic motion, we would have $h=m$, where $m$ is the baryonic mass, and ($C_2,\,C_3$) would reduce to ($mu_{\mu}\xi_t^{\mu},\,mu_{\mu}\xi_{\phi}^{\mu}$) that are the usual energy and angular momentum conservation laws, respectively. The constants ($hu_{\mu}\xi_t^{\mu},\,hu_{\mu}\xi_{\phi}^{\mu}$) are their generalizations to the case where the fluid is subject to acceleration, which takes place when the pressure throughout the fluid is not uniform.

Since the transverse three-velocity is subject to $u^2<1$~\eqref{0m7}, this restricts the values of $\Om$ by
\begin{multline}\label{0t5}
\Om_-<\Om <\Om_+,
\\ \Om_{\pm}\equiv \om \pm \sqrt{\om^2-\tfrac{g_{tt}}{g_{\phi\phi}}}=\om\pm \tfrac{\sqrt{D}}{|g_{\phi\phi}|}.
\end{multline}
It is easy to show that $\Om_-$ is greater than $-g_{tt}/g_{t\phi}$ ensuring $C_2<0$~\eqref{t4} for all\footnote{The sign of $C_2$ is related to the metric signature ($-,\,+,\,+,\,+$). Had we chosen the signature ($+,\,-,\,-,\,-$) for~\eqref{0m1} we would have $C_2>0$. What counts is that $C_2\neq 0$.} $r$. It is also easy to show that $\Om_-<0,\ \Om_+>2\om >0$ outside the ergoregion, $\Om_-=0,\ \Om_+=2\om$ on the boundary of the ergoregion, and $\Om_->0,\ \om<\Om_+<2\om$ inside the ergoregion. Since $D_h=0$, we have $\Om_-(r_h)=\Om_+(r_h)=\om(r_h)$.

On combining~\eqref{t4} and~\eqref{t5} we obtain, setting $C_4\equiv C_3/C_2$,
\begin{equation}\label{t9}
    \Om(r) =\frac{\om+C_4\tfrac{g_{tt}}{g_{\phi\phi}}}{1+C_4\om}.
\end{equation}
The value $C_4=0$ yields $\Om =\om$ corresponding to the case where the fluid is only being dragged by the rotating solution. This results in $u\equiv 0$~\eqref{0m7}. We see that the fluid flow is characterized by three angular velocities: the one equals the angular velocity of the ZAMO's, $\om(r)$, corresponding to $C_4=0$ and the other two are $\Om_-(r)$ and $\Om_+(r)$ corresponding to the maximum value and minimum value of $C_4$ if $r\neq r_h$, respectively. Note that the derivative of the rhs in~\eqref{t9} with respect to $C_4$, $-D/[g^2_{\phi\phi}(1+C_4\om)^2]$, is negative for $r\neq r_h$ and vanishes for $r=r_h$. The three angular velocities are equal on the horizons.

If the solution is static (nonrotating with $\om\equiv 0$), it is apparent from~\eqref{t9} that the angular velocity of the fluid elements is function of the ratio $g_{tt}/g_{\phi\phi}$
\begin{equation}\label{t10}
    \Om_{\text{stat}}(r)=C_4\frac{g_{tt}}{g_{\phi\phi}}.
\end{equation}

On inserting~\eqref{g2} into~\eqref{g1b}, \eqref{t4} and~\eqref{t5} we obtain
\begin{align}\label{t7}
&(\text{a})\;\frac{|g_{\ta\ta}|Dn^2v^2}{(1-V^2)}=C_1^2,\nn\\
&(\text{b})\;\frac{h^2(g_{tt}+\Om g_{t\phi})^2|g_{\phi\phi}|}{D(1-V^2)}=\frac{-h^2(g_{tt}+\Om g_{t\phi})^2}{(g_{tt}+\om g_{t\phi})(1-V^2)}=C_2^2,\nn\\
&(\text{c})\;\frac{h^2(\Om -\om)^2|g_{\phi\phi}|^3}{D(1-V^2)}=\frac{h^2u^2|g_{\phi\phi}|}{1-V^2}=C_3^2,
\end{align}
where we have used another expression for $D=-|g_{\phi\phi}|(g_{tt}+\om g_{t\phi})$ and~\eqref{0m7}. Notice that the case $C_1=0$ corresponds either to $v\equiv 0$ (circular flow) or to $n\equiv 0$ (no fluid). We drop this case from our analysis and focus on the case $C_1^2>0$.

\section{End-behavior flow\label{seceb}}
In this section we consider the fluid flow in the vicinity of the horizons and the flow as the fluid extends to spatial infinity.

\subsection{Flow in the vicinity of the horizons\label{secv}}
In equations~\eqref{t4} and~\eqref{t5}, $u^t$ is given by~\eqref{g2} and it diverges by~\eqref{ta3} as the fluid approaches a horizon. One of the factors in each of Eq.~\eqref{t4} and Eq.~\eqref{t5} has to go to zero to keep the value of the lhs constant. For ordinary matter, $h$ has a lower limit $m$, the only option is to have in~\eqref{t4} $\Om (r_h)=-g_{tt}(r_h)/g_{t\phi}(r_h)$, which is by~\eqref{0m2} and~\eqref{ta3}
\begin{equation}\label{t6}
    \Om_h=-\frac{g_{t\phi}(r_h)}{g_{\phi\phi}(r_h)}=\om_h.
\end{equation}
This is the same equation derived from~\eqref{t5}. This results in $u_h=0$~\eqref{0m7} and by~\eqref{m7} in
\begin{equation}\label{t6b}
V_h^2=v_h^2.
\end{equation}
The fluid is doomed to rotate at ZAMO's angular velocity as it approaches a horizon. Particularly, if the black hole is static, the fluid reaches a horizon with a vanishing angular velocity.

Based on our previous conclusions, $D_h=0$ and $\Om_h=\om_h$, it is easy to show that the fluid reaches, or emanates from, a horizon with a three-dimensional radial velocity $|v_h|= 0$ (subsonic velocity on the horizons) or $|v_h|= 1$ (luminal velocity on the horizons)~\cite{add}. In fact, since $D_h=0$ and $V_h^2=v_h^2$, equation~(\ref{t7}-a) would imply $C_1=0$ if $0<v_h<1$, which is the case we dropped from our analysis and we only considered the case $C_1>0$. In the former case, $|v_h|= 0$, the particle density $n_h\to\infty$. Note that this conclusion, $|v_h|= 0$ or $|v_h|= 1$, is solely based on the continuity equation~\eqref{g1} and the assumptions made in the first two paragraphs of Sec.~\ref{assump}. This conclusion applies to non-perfect fluids too provided we adopt the so-called Eckart frame or particle frame for defining the four-velocity vector of the fluid~\cite{Rezzolla,Ellis}. In this frame, $J^{\mu}$ and $u^{\mu}$ are parallel and~\eqref{g1} is still valid.

The case $|v_h|= 1$ is physically more interesting. The case $|v_h|= 0$ has been discussed in the literature too~\cite{acc2,aafj,aafs,add,Kafka,Gillman}. In Ref.~\cite{acc2} the cases with subsonic velocities at the horizon~\cite{Kafka,Gillman} were described as unphysical and in Refs.~\cite{aafj,aafs,add} it has been shown that if $v_h\to 0$ then the pressure increasingly diverges as the fluid approaches a horizon yielding a divergent gradient of pressure opposing the flow motion and pushing the fluid backward. There are no available empirical data to support such a conclusion nor are there data to refute it. In this work we too drop this case from our analysis and focus only on the case $|v_h|= 1$.

Now back to Eqs.~(\ref{t7}-a) and~(\ref{t7}-c). In the limit $r\to r_h$, we have respectively
\begin{align}
\label{t7li1}&n^2=\frac{C_1^2(1-v^2-u^2)}{|g_{\ta\ta}|Dv^2} \underset{(r\to r_h)}{\propto} \frac{1-v^2-u^2}{D},\\
\label{t7li2}&h^2=\frac{C_3^2(1-v^2-u^2)}{|g_{\phi\phi}|u^2} \underset{(r\to r_h)}{\propto} \frac{1-v^2-u^2}{u^2},
\end{align}
For most known nonextremal black holes $D\sim |r-r_h|$ as $r\to r_h$. To include in the investigation other types of black holes, including phantom~\cite{GR,phantom,multi} and regular~\cite{regular1,reg,gen2} black holes, we consider the case
\begin{equation}\label{t8li}
D \underset{(r\to r_h)}{\simeq} C_D|r-r_h|^x\qquad\quad (x>0),
\end{equation}
where $x$ is a measure of the horizon degeneracy. It is understood that $r$ approaches $r_h$ from within the region of the three-space where the Killing vector $\xi_t^{\mu}=(1,\,0,\,0,\,0)$ is timelike.

In the vicinity of a horizon we set
\begin{equation*}
1-v^2\simeq C_v|r-r_h|^y,\qquad u^2\simeq C_u^2|r-r_h|^{z},
\end{equation*}
with $z\geq y$ to ensure that $h^2>0$~\eqref{t7li2}; if $z=y$ we further restrict $C_v$ and $C_u$ by $C_v/C_u^2>1$ to have $h^2>0$. This yields
\begin{align}\label{t9li}
&v \underset{(r\to r_h)}{\simeq} 1-C_v\tfrac{|r-r_h|^y}{2},\qquad u \underset{(r\to r_h)}{\simeq} C_u|r-r_h|^{z/2},\nn\\
&n \underset{(r\to r_h)}{\simeq} C_n|r-r_h|^{(y-x)/2},\\
&h \underset{(r\to r_h)}{\simeq}\left\{
                                \begin{array}{ll}
                                  C_h|r-r_h|^{(y-z)/2}, & \hbox{$z>y$;} \\
                                  C_h, & \hbox{$z=y$ \text{ and } $C_v/C_u^2> 1$,}
                                \end{array}
                              \right.\nn
\end{align}
where $C_v>0$, $C_u>0$, $C_n>0$, and $C_h>0$ are finite constant values ($C_h=h_h$ if $z=y$ and $C_v/C_u^2> 1$). The third line in~\eqref{t9li}, corresponding to $z>y$, has been derived by first rewriting~\eqref{t7li2} as
\begin{equation*}
    h^2\underset{(r\to r_h)}{\propto} \frac{1-v^2}{u^2}-1.
\end{equation*}
For $z>y$, the first term diverges in the limit $r\to r_h$, so we can drop 1 and have $h\underset{(r\to r_h)}{\propto}\sqrt{1-v^2}/u$ yielding the third line in~\eqref{t9li}.

Using the law of motion~\cite{Gravitation,Rezzolla}
\begin{equation}\label{lm}
    nha^{\mu}=-(g^{\mu\nu}+u^{\mu}u^{\nu})\nabla_{\nu}p,
\end{equation}
where $a^{\nu}\equiv u^{\mu}\nabla_{\mu}u^{\nu}$ is the four-acceleration vector, we arrive at (see Appendix)
\begin{equation}\label{t10li}
\partial_{r}p\underset{(r\to r_h)}{\sim} |r-r_h|^{y-1-(x+z)/2}.
\end{equation}
Two physical cases emerge depending on the value of $y-1-(x+z)/2$. For a reason that will become clear in Sec.~\ref{secndd}, we impose the ``physical" constraint that the pressure has no logarithmic divergence as the fluid approaches a horizon
\begin{equation}\label{t11li}
   y-1-\tfrac{(x+z)}{2}\gtrless -1\qquad \big(y-\tfrac{(x+z)}{2}\gtrless 0\big),
\end{equation}
where the upper sign yields a finite pressure and the lower sign a divergent pressure as $r\to r_h$. In both cases the pressure behaves generically as
\begin{equation}\label{t12lib}
 p\underset{(r\to r_h)}{\simeq} p_h+C_p|r-r_h|^{y-(x+z)/2},
\end{equation}
where $C_p$ and $p_h$ are constants. By a generic behavior we mean that the constraints~\eqref{t11li} do not depend on the metric value on the horizons; they just depend on the parameters ($x,\,y,\,z$). In a non-generic behavior it would be possible, as shown in the last paragraph of the Appendix, to fix $C_p=0$ by imposing constraints relating ($x,\,y,\,z$) to the metric value on the horizons. In that case the pressure would behave as $p\underset{(r\to r_h)}{\simeq} p_h+\bar{C}_p|r-r_h|^{1+y-(x+z)/2}$ where $\bar{C}_p$ is a constant, provided $y-(x+z)/2\neq -1$. To keep the discussion as generic as possible we drop this case from our consideration.

Two physical cases emerge from the above discussion:
\begin{enumerate}
  \item \textsl{Finite pressure in the vicinity of the horizons}. This case with the upper sign in~\eqref{t11li} yields, using the fact that $z\geq y$,
\begin{equation}\label{t12li}
  y-x>z-y\geq 0.
\end{equation}
This implies that $n\to 0$~\eqref{t9li} as $r\to r_h$. We distinguish two subcases:
\begin{description}
  \item[] (a) Non-geodesic flow in the vicinity of the horizons - This corresponds to
        \begin{equation}\label{t13li}
        0\geq y-1-\tfrac{(x+z)}{2}>-1.
        \end{equation}
        Since in this case the pressure gradient~\eqref{t10li} is nonzero, the flow is non-geodesic in the vicinity of the horizons.
  \item[] (b) Geodesic flow in the vicinity of the horizons - This corresponds to
        \begin{equation}\label{t14li}
        y-1-\tfrac{(x+z)}{2}>0.
        \end{equation}
        In this case $\partial_{r}p\to 0$ as $r\to r_h$ resulting in a geodesic flow in the vicinity of the horizons. Note that the pressure~\eqref{t12lib} falls to a constant value, as $r\to r_h$, in a way faster than in the previous subcase.
\end{description}
  \item \textsl{Divergent pressure in the vicinity of the horizons}. This case with the lower sign in~\eqref{t11li} yields
        \begin{equation}\label{t13lib}
        y-x<z-y.
        \end{equation}
        This, for instance, may be satisfied if we take $z>y$ and $x>y$ or $z=y$ and $x>y$, in which cases the number density $n$ diverges as $r\to r_h$.
\end{enumerate}

\subsection{Asymptotic Flow\label{secaf}}
If the fluid flow may extend to spatial infinity (this concerns solutions with no cosmological horizon), the flow is said to be global.

For stars $\om\sim 1/r^3$ as $r\to\infty$. If the metric~\eqref{0m1} is asymptotically flat (as. flat), $g_{tt}\sim 1$ and $g_{\phi\phi}\sim r^2$ as $r\to\infty$, and if the flow is global, then by~\eqref{t9} we have $\Om\sim 1/r^2$ as $r\to\infty$. This applies to $\Om_{\text{stat}}$~\eqref{t10} too. If the metric~\eqref{0m1} is asymptotically anti-de Sitter (as. adS), $g_{tt}\sim r^2$ and $g_{\phi\phi}\sim r^2$ as $r\to\infty$, and if the fluid flow is global, we have that $\Om\sim \mathrm{const}$ and $\Om_{\text{stat}}\sim \mathrm{const}$ as $r\to\infty$. Using this in~(\ref{t7}-a) along with $g_{t\phi}\sim 1/r$, $D\sim r^2$ (as. flat), $D\sim r^4$ (as. adS)~\eqref{0m2}, and $u\sim 1/r^2$ (as. flat), $u\sim 1/r^4$ (as. adS)~\eqref{0m7} as $r\to\infty$, we obtain
\begin{equation}\label{af1}
\frac{n^2v^2}{1-v^2}\sim \frac{C_1^2}{r^4}\;\ \text{(as. flat)},\qquad \sim \frac{C_1^2}{r^6}\;\ \text{(as. adS)}.
\end{equation}
On performing similar end-behavior evaluations on~(\ref{t7}-b) we arrive at
\begin{equation}\label{af2}
\frac{h^2}{1-v^2}\sim C_2^2\;\ \text{(as. flat)},\qquad \sim \frac{C_2^2}{r^2}\;\ \text{(as. adS)}.
\end{equation}
In all cases, we have
\begin{equation}\label{af3}
\frac{n^2v^2}{h^2}\sim \frac{1}{r^4}\qquad (\text{as }r\to\infty).
\end{equation}

\section{Applications: Non-degenerate and degenerate gases\label{secndd}}
What distinguishes a black hole from a star is the presence of horizons in the vicinity of which interesting physical phenomena may occur due to strong gravitational attraction. It is worth mentioning that the Hawking radiation occurs there. Another well instance and established fact concerns the efficiency of emission due to accreting matter onto a black hole that is few orders smaller than that due to accreting matter onto a neutron star~\cite{Kremer}. These two instances show that the investigation of physical phenomena in the vicinity of horizons is of great importance.

In this section we focus more on accretion, as well as particle jets, in the vicinity of horizons. We will show that the accretion model we have developed in the previous sections applies to a variety of non-degenerate and degenerate gases. The applications mostly concern ultra- and non-relativistic flows (at least in the vicinity of the horizons) where the ansatz~\eqref{t9li}$_1$ holds\footnote{By the notation $(\text{i})_\text{j}$ we mean the j-th line in equation (i).}.

The statistics of non-degenerate and degenerate gases result in different one-particle distribution functions~\cite{Rezzolla,Muller,Kremer}. Non-degenerate gases obey Maxwell-Boltzmann statistics and degenerate gases obey either Fermi-Dirac statistics, if they have half-integral spin, or Bose-Einstein statistics, if they have integral spin.

We are not interested in the equilibrium distribution function corresponding to each of the three cases mentioned above; rather, we are concerned with the expressions of the fields in equilibrium, which are the number and energy densities ($n,\,\epsilon$), enthalpy $h$ and entropy $s$ per particle, pressure $p$, and chemical potential $\mu$, and we seek to apply them to the regions adjacent to the horizons where the Killing vector $\xi_t^{\mu}=(1,\,0,\,0,\,0)$ is timelike.

Applications of equilibrium distribution functions are well-known in the literature~\cite{app1} and include application to white dwarf stars~\cite{app2}. We intend to extend their realm to accretion onto black holes.

It is according to the limiting values of the dimensionless parameter $\zeta\equiv mc^2/k_{\text{B}}T$ (coldness), where $T$ is the temperature of the fluid and $k_{\text{B}}$ is the Boltzmann constant, that one defines the ultra-relativistic ($\zeta\ll1$) and non-relativistic ($\zeta\gg1$) cases. So, the ultra-relativistic limit corresponds generally to high temperatures and the non-relativistic limit corresponds generally to low temperatures. It has become customary to let the speed of light $c$ appear explicitly in $\zeta$ and the expressions of the fields of this section. As to the states of non-degeneracy ($\alpha_{\text{f}}\ll 1$) and degeneracy ($\alpha_{\text{f}}\gg 1$) of the fluid are defined by the limiting values of the fugacity $\alpha_{\text{f}}\equiv \mu/k_{\text{B}}T$, which is a dimensionless parameter expressing the tendency of volatility~\cite{Huang}.

\subsection{Non-degenerate fluids: $\alpha_{\text{f}}\ll 1$\label{secnd}}

In terms of the modified Bessel function of second kind $K_{\nu}$~\cite{hand}, the expressions of ($n,\,p,\,h,\epsilon$) are given by (see, for instance, \cite{Rezzolla,Muller,Kremer})
\begin{align}
\label{app1}&n=4\pi m^2ck_{\text{B}}T\frac{g_{\text{s}}}{h^3_{\text{P}}}K_2(\zeta)\e^{\mu/k_{\text{B}}T},& & p=nk_{\text{B}}T,\\
\label{app2}&h=mc^2\frac{K_3(\zeta)}{K_2(\zeta)},& &  \epsilon =nh-p,
\end{align}
where $h_{\text{P}}$ is the Planck constant.

\subsubsection{Ultra-relativistic limit: $\zeta\ll1$}
The ultra-relativistic limit of $h$ is to the leading terms ($K_2\simeq 2\zeta^{-2}$, $K_3\simeq 8\zeta^{-3}$) given by
\begin{equation}
h\simeq 4k_{\text{B}}T,
\end{equation}
which is large and remains finite. This corresponds to the case $z=y$~\eqref{t9li}. Non-degenerate fluids are subject to $\alpha_{\text{f}}\ll 1$ resulting in $\e^{\mu/k_{\text{B}}T}\ll 1$ and $n\ll 1$~\eqref{app1}. The pressure, proportional to $n$~\eqref{app1}, goes to zero as $n$ does. This corresponds to $y>x$~\eqref{t9li} and $p_h=0$~\eqref{t12lib} so that both $p$ and $n$ behave as $|r-r_h|^{(y-x)/2}$ in the limit $r\to r_h$. Using the above expressions we obtain $p\simeq \epsilon/3$, which implies that $\epsilon\to 0$ as $r\to r_h$; the energy per particle, however, does not go to zero as $r\to r_h$ and it is given by $\epsilon/n\simeq 3k_{\text{B}}T$.

\subsubsection{Non-relativistic limit: $\zeta\gg1$}
In this limit we have
\begin{equation}
h\simeq mc^2+5k_{\text{B}}T/2,
\end{equation}
and this again yields $z=y$, $h_h=mc^2$, $y>x$~\eqref{t9li} and $p_h=0$~\eqref{t12lib} so that both $p$ and $n$ behave as $|r-r_h|^{(y-x)/2}$ in the limit $r\to r_h$.

We have thus shown that ultra-relativistic, as well as non-relativistic, non-degenerate fluids reach horizons in a state of a very dilute matter and vanishing pressure and energy density as measured by a local Lorentz rest frame that moves with the fluid velocity. Equation~\eqref{app1} shows that the chemical potential diverges logarithmically as $r\to r_h$. This effect mimics that of the center-of-mass-energy (CME) of two colliding particles in the vicinity of a horizon by which the CME attains an arbitrarily large value~\cite{1}. This may signal the necessity of introducing quantum effects in the vicinity of the horizons.

Recall that the chemical potential measures the change in the internal energy (the energy in the fluid's rest frame) as one particle is added to the system. Since the chemical potential becomes arbitrarily large and negative as the fluid approaches a horizon, the addition of one particle results in an arbitrarily large decrease in the internal energy. Conversely the extraction of one particle would require an increasingly large amount of energy as the fluid approaches a horizon. Thus, the black hole ensures cohesion and stability of the accreting matter.

\subsection{Completely degenerate Fermi fluids: $\alpha_{\text{f}}\gg 1$\label{secd1}}
The ``exact" expressions of ($n,\,p,\,\epsilon$) for the Fermi-Dirac fluids, as for the Bose-Einstein fluids, are available in the literature (see, for instance, \cite{Rezzolla,Muller,Kremer,Huang}). We content here to give only their ultra- and non-relativistic limits.

\subsubsection{Ultra-relativistic limit: $p_{\text{F}}\gg mc$}
Fermi fluids are characterized by the presence of a limiting momentum value called Fermi momentum $p_{\text{F}}$ all the quantum states with $|\vec{p}|<p_{\text{F}}$ are filled by fermions and the states with $|\vec{p}|>p_{\text{F}}$ are empty. Since the fields do not depend (explicitly) on the temperature the coldness parameter $\zeta$ is no longer convenient for discussing the limiting ultra- and non-relativistic cases. These two limits correspond to $p_{\text{F}}\gg mc$ and $p_{\text{F}}\ll mc$, respectively.

Rather, the fields ($\epsilon,\,p,\,h$) depend only on $n$. To the leading term in $n$, which is supposed to be large enough to allow for series expansions in powers of $1/n$, the expressions of ($\epsilon,\,p$) read
\begin{equation}
p\simeq \frac{\epsilon}{3},\qquad p\simeq \frac{1}{4}\Big(\frac{3c^3h^3_{\text{P}}}{4\pi g_{\text{s}}}\Big)^{1/3}n^{4/3}=\frac{\pi g_{\text{s}}c}{3h^3_{\text{P}}}p^4_{\text{F}},
\end{equation}
from which we derive
\begin{equation}\label{app3}
h\simeq \Big(\frac{3c^3h^3_{\text{P}}}{4\pi g_{\text{s}}}\Big)^{1/3}n^{1/3},\qquad h\simeq cp_{\text{F}}.
\end{equation}
This corresponds to the case
\begin{equation}
z>y\quad \text{ and }\quad x>y,
\end{equation}
where the pressure~\eqref{t13lib}, the number density and enthalpy per particle~\eqref{t9li} all attain arbitrarily large values in the vicinity of the horizons. The Fermi momentum $p^3_{\text{F}}=3h^3_{\text{P}}n/(4\pi g_{\text{s}})$ also attains an arbitrarily large value. This effect mimics the classical CME effect and may be ruled out, as well as the CME effect, by the introduction of quantum corrections.

From~\eqref{t9li} and the first equation in~\eqref{app3} we derive
\begin{equation}\label{app4}
x=3z-2y.
\end{equation}
Since $z>y$, this implies $x>z$. Thus, as the flow of the ultra-relativistic Fermi fluid approaches a horizon, the parameters ($y,\,z$) adjust their values to remain smaller than $x$ ($x>z>y$) while subject to the constraint~\eqref{app4}.

\subsubsection{Non-relativistic limit: $p_{\text{F}}\ll mc$}
In this limit $n$ is supposed to be small enough to allow for series expansions in powers of $n$. To the leading order of approximation the expression of ($\epsilon,\,p$) read
\begin{multline}
\epsilon\simeq nmc^2+\frac{1}{10}\Big(\frac{3h^3_{\text{P}}}{4\pi m^{3/2}g_{\text{s}}}\Big)^{2/3}n^{5/3}=nmc^2+\frac{2\pi g_{\text{s}}}{15mh^3_{\text{P}}}p^5_{\text{F}},\\ p\simeq \frac{2(\epsilon-nmc^2)}{3}=\frac{4\pi g_{\text{s}}}{45mh^3_{\text{P}}}p^5_{\text{F}},
\end{multline}
from which we derive
\begin{equation}\label{app5}
h\simeq mc^2+\frac{1}{2}\Big(\frac{3h^3_{\text{P}}}{4\pi m^{3/2}g_{\text{s}}}\Big)^{2/3}n^{2/3}=mc^2+\frac{1}{2m}p^2_{\text{F}}.
\end{equation}
This corresponds to the case $z=y$, $y>x$, $h_h=mc^2$~\eqref{t9li}, and $p_h=0$~\eqref{t12lib}. We see that the non-relativistic Fermi fluid behaves in the vicinity of the horizons as does the non-relativistic, non-degenerate fluid behave with vanishing pressure and energy density. Moreover, the Fermi momentum, $p^3_{\text{F}}=3h^3_{\text{P}}n/(4\pi g_{\text{s}})$, attains an arbitrarily small value.

The progressive vanishing of the Fermi momentum as $r\to r_h$ has the following physical interpretation. All the quantum states filled with fermions, having $|\vec{p}|<p_{\text{F}}$, form progressively a continuum of energy as $r\to r_h$, that is, as $p_{\text{F}}\to 0$. The momentum particle density\footnote{In contrast with the volume particle density $n=\dd N/\dd x^1\dd x^2\dd x^3\propto p^{3}_{\text{F}}$ with $\sqrt{\sum_{i=1}^3(x^i)^2}=r$.} $\dd N/\dd p^1\dd p^2\dd p^3\propto p^{-3}_{\text{F}}$, with $\sqrt{\sum_{i=1}^3(p^i)^2}=|\vec{p}|$ and $N$ being the fermions number, assumes an arbitrarily large value in the vicinity of the horizons.

\subsection{Degenerate Bose fluids: $\alpha_{\text{f}}\gg 1$\label{secd2}}
Here again we give only the limiting expressions of ($n,\,p,\,\epsilon$) for the Bose-Einstein fluids (see, for instance, \cite{Rezzolla,Muller,Kremer,Huang} for more details).

The Bose-Einstein condensation occurs for the ultra- and non-relativistic limits at the well-known temperatures
\begin{multline}\label{app6}
  T_{\text{cond}}=\frac{hc}{2k_{\text{B}}}\Big[\frac{n}{\pi g_{\text{s}}\zeta_R(3)}\Big]^{1/3},\\ T_{\text{cond}}=\frac{h^2}{2\pi mk_{\text{B}}}\Big[\frac{n}{g_{\text{s}}\zeta_R(3/2)}\Big]^{2/3},
\end{multline}
respectively. Here $\zeta_R(i)=\sum_{j=1}^{\infty}j^{-i}$ is the Riemann zeta function. Condensation means that some fraction $n_{\text{cond}}$ of $n$ occupies the level with zero energy. This fraction is almost zero for $T>T_{\text{cond}}$ ($n_{\text{cond}}\ll n$) and becomes important for $T<T_{\text{cond}}$ according to the laws
\begin{equation}
 \frac{n_{\text{cond}}}{n}=1-\Big[\frac{T}{T_{\text{cond}}}\Big]^{3},\quad \frac{n_{\text{cond}}}{n}=1-\Big[\frac{T}{T_{\text{cond}}}\Big]^{3/2},
\end{equation}
for the ultra- and non-relativistic limits, respectively. The pressure obeys the same laws as for Fermi fluids
\begin{equation}\label{app7}
p\simeq \frac{\epsilon}{3},\qquad p\simeq \frac{2(\epsilon-nmc^2)}{3},
\end{equation}
for the ultra- and non-relativistic limits, respectively.

We assume that initially $T<T_{\text{cond}}$.

\subsubsection{Ultra-relativistic limit of the condensation: $\zeta\ll1$}
In this limit the pressure is proportional to the power four of the temperature
\begin{equation}\label{app8}
p\simeq 8\pi\zeta_R(4)\Big(\frac{k_{\text{B}}}{h_{\text{P}}c}\Big)^3g_{\text{s}}k_{\text{B}}T^4.
\end{equation}
So the pressure remains finite implying the inequalities~\eqref{t12li}. By~\eqref{t9li} $n$ goes to zero as the Bose fluid approaches a horizon and thus, by~\eqref{app6}, $T_{\text{cond}}\to 0$ as $r\to r_h$. There are two possible scenarios: `$T$ constant' scenario and `$T$ non-constant' scenario.\\

\paragraph{$T$ constant: Isothermal flow.} Since  $T_{\text{cond}}\to 0$ as $r\to r_h$, once $T_{\text{cond}}$ becomes smaller than $T$ the condensation ceases, then completely disappears, as the ultra-relativistic Bose fluid approaches a horizon.

Using~\eqref{ent} and~\eqref{app7} we see that $h\simeq 4p/n=4p_h|r-r_h|^{-(y-x)/2}/C_n$ where $p_h$, defined in~\eqref{t12lib}, is the rhs of~\eqref{app8}. So $h$ diverges in the limit $r\to r_h$. On comparing this expression of $h$ with that in~\eqref{t9li} we obtain
\begin{equation}\label{app9}
    x=2y-z\qquad (z>y>x).
\end{equation}

\paragraph{$T$ non-constant.}
If the fluid heats during accretion the condensation ceases as in the `$T$ constant' scenario. If, instead, $T$ decreases and remains smaller than $T_{\text{cond}}$, $T$ must go to zero too in the limit $r\to r_h$. While the ultra-relativistic approximation does not hold in the limit $T\to 0$ we can still conclude that as an ultra-relativistic Bose fluid approaches a horizon its temperature gradually goes to zero if initially $T<T_{\text{cond}}$; in the very vicinity of a horizon, as the state of the fluid evolves from ultra- to non-relativistic, one has to use the exact expressions of the fields, for the ultra-relativistic approximation is no longer valid for $T\to 0$.

From~\eqref{app6} we see that $n$ does not depend on $T$; however, since both $n$ and $T$ decrease as the Bose fluid approaches a horizon, $n$ tends to become an implicit function of $T$. In this case, as the fluid evolves from the ultra-relativistic state to the non-relativistic state, $h\simeq 4p/n$ converges to $mc^2$ in the limit $r\to r_h$ and this yields $z=y>x$.

\subsubsection{Non-relativistic limit of the condensation: $\zeta\gg1$}
The pressure and the enthalpy are given by
\begin{equation}
p\simeq \zeta_R(5/2)\Big(\frac{2\pi mk_{\text{B}}}{h^2_{\text{P}}c}\Big)^{3/2}g_{\text{s}}k_{\text{B}}T^{5/2},
\end{equation}
$h\simeq mc^2+5p/(2n)$. The two scenarios described in the ultra-relativistic limit occur in this limit too.

In the `$T$ constant' scenario the same conclusions and equation~\eqref{app9} remain valid.

In the `$T$ non-constant' scenario the non-relativistic approximation remains valid for all $T$ down to zero. In this approximation and the limit $T\to 0$, the enthalpy $h$ converges to $mc^2$. This yields $z=y>x$.

It is now clear that a logarithmic divergence in the expression of the pressure~\eqref{t11li} would not be supported by any of the statistical models described above. Moreover, with the exception of the ultra-relativistic limit of the completely degenerate Fermi fluids where the pressure attains arbitrarily large values, by the results derived above for the other statistical cases, the pressure either converges to a constant value or vanishes in the limit $r\to r_h$.

\section{Thermodynamics and critical flow\label{seccf}}
All perfect fluids obey the adiabatic conservation law~\cite{Gravitation} $u^{\mu}\nabla_{\mu}s=0$: The entropy per particle is conserved along the fluidlines. By the assumptions made in the first two paragraphs of Sec.~\ref{assump}, this results in $s=\mathrm{const}$ throughout the accreting planar disk, a property by which the fluid is said to be isentropic.

From the two thermodynamic laws~\cite{Rezzolla,Gravitation}, $\dd p=n(\dd h-T\dd s)$ and $\dd\ep=h\dd n+nT\dd s$ ($T$ being the temperature) applied to the accreting disk with $\dd s=0$, we obtain the following equation
\begin{equation}\label{cf0}
\frac{\dd h}{h}=\al^2~\frac{\dd n}{n},
\end{equation}
where $\al$, such that $\al^2\equiv\dd p/\dd\ep$, is the isentropic (also adiabatic) three-dimensional sound speed.

The critical behavior of the fluid corresponds to the stationary values of the functions~\eqref{t7}. There are many ways to choose the dynamical variables~\cite{aafj}. Following the latter reference, we choose ($r,\,v,\,\Om$) as variables for the dynamical system. Since the constant of motion $C_2$ has the dimension of energy, it will be more appropriate to look for the stationary values of the function in~(\ref{t7}-b) which represents the energy squared of a fluid particle. We denote it by $H$:
\begin{equation}\label{cf1}
    H(r,v,\Om)=\frac{-h^2(g_{tt}+\Om g_{t\phi})^2}{(g_{tt}+\om g_{t\phi})(1-V^2)}.
\end{equation}
The stationary values of $H$ are subject to the constraint~(\ref{t7}-a) and~(\ref{t7}-c).

Let ($F_r,\,F_v,\,F_{\Om}$) denote the partial derivatives of $F$ with respect to ($r,\,v,\,\Om$), respectively, where $F$ is some function of ($r,\,v,\,\Om$) and let $F_i$ represents any of ($F_r,\,F_v,\,F_{\Om}$). We intend to determine the critical points (CPs) which we denote by ($r_c,\,v_c,\,\Om_c$). These are the points solutions to $H_r=0$, $H_v=0$, and $H_{\Om}=0$. From~(\ref{t7}-a) and $(\ln h)_i=\al^2(\ln n)_i$~(\ref{cf0}) we obtain
\begin{align}\label{cf2}
&(\ln h)_r=-\tfrac{\al^2}{2}\Big(\tfrac{2uu_r}{1-V^2}+\ln(|g_{\ta\ta}|D)_r\Big),\nn\\
&(\ln h)_v=-\al^2~\tfrac{1-u^2}{v(1-V^2)},\\
&(\ln h)_{\Om}=-\al^2~\tfrac{g_{\phi\phi}^2(\Om-\om)}{D(1-V^2)},\nn
\end{align}
where $u$ is given by~\eqref{0m7}. Direct calculations reveal
\begin{align}
\label{cf3}&H_{\Om}=+\frac{2h^2|g_{\phi\phi}|(g_{tt}+\Om g_{t\phi})f_1(r,v,\Om)}{(g_{tt}+\om g_{t\phi})^2(1-V^2)^2},\\
\label{cf4}&H_{v}=+\frac{2h^2(g_{tt}+\Om g_{t\phi})^2f_2(r,v,\Om)}{v(g_{tt}+\om g_{t\phi})^2(1-V^2)^2},
\end{align}
where
\begin{align*}
&f_1=[(1-\al^2)\Om-(v^2-\al^2)\om]g_{tt}\\
&\quad +[\al^2\Om^2-(1+\al^2)\Om\om+v^2\om^2]g_{\phi\phi},\\
&f_2=(v^2-\al^2)g_{tt}-(\al^2\Om^2-2\al^2\Om\om+v^2\om^2)g_{\phi\phi}.
\end{align*}
In the discussion following Eq.~\eqref{0t5} we have shown that $g_{tt}+\Om g_{t\phi}>0$ and hence the equations $H_{\Om}=0$ and $H_{v}=0$ yield $f_1=0$ and $f_2=0$. Evaluating $f_1+\om f_2=0$ we obtain $\Om (1-\al^2)D/g_{\phi\phi}=0$, which yields the first critical value
\begin{equation}\label{cf5}
    \Om_c=0.
\end{equation}
This shows that if initially the fluid is not rotating ($\Om_{\text{init}}=0$) or if it corotates with the black hole ($\Om_{\text{init}}>0$), there will be no critical flow, for the black hole drags the fluid so that $\Om >0$ for the the whole accretion time. The critical flow occurs only if initially the fluid rotates retrograde ($\Om_{\text{init}}<0$). If the black hole is static then the critical flow occurs if initially the fluid is not rotating~\cite{aafj,aafs,add}.

Since $D_h=0$~\eqref{ta3}, $r=r_h$ is another solution to $f_1+\om f_2=0$ and it may provide another critical value: This is rather an end-point and not a CP in the mathematical sense.

Substituting this critical value $\Om=\Om_c$ into the expression of $f_1=0$ (or $f_2=0$), we obtain
\begin{equation}\label{cf6}
    v_c^2=\frac{g_{tt\,c}}{g_{tt\,c}-g_{\phi\phi\,c}\om_c^2}~\al^2=-\frac{g_{tt\,c}g_{\phi\phi\,c}}{D_c}~\al^2<\al^2,
\end{equation}
where $g_{tt\,c}=g_{tt}(r_c)$, $g_{\phi\phi\,c}=g_{\phi\phi}(r_c)$, $\om_c=\om(r_c)$, and $D_c=D(r_c)$. We see that at the CP the radial three-velocity is different from the fluid's adiabatic sound speed. Only in the static case, $\om\equiv 0$, we do have $v_c^2=\al^2$. Thus, the conclusions and results of the static case do not generalize to the rotating configurations. The three-speed $V_c^2$ is also different from $\al^2$ and reduces to it in the static limit ($\om\equiv 0$) only. This is evaluated from~\eqref{0m7} and~\eqref{cf6} as
\begin{multline}\label{cf7}
    1>V_c^2=\tfrac{g_{t\phi\,c}^2-\al^2g_{tt\,c}g_{\phi\phi\,c}}{D_c}=1+\tfrac{g_{tt\,c}g_{\phi\phi\,c}}{D_c}~(1-\al^2)\\
    =\al^2+\tfrac{g_{t\phi\,c}}{D_c}~(1-\al^2)>\al^2.
\end{multline}

The unconstrained expression of $H_r$ is sizeable. However, using the already determined extreme values~\eqref{cf5} and~\eqref{cf7} to eliminate ($\Om,\,V^2$), this reduces after some algebra to
\begin{equation}\label{cf8}
\hspace{-1.89mm}H_r=+\frac{h^2g_{tt}}{1-\al^2}~[(\ln |g_{tt}|)_r-\al^2(\ln|g_{\ta\ta}|D)_r]\bigg|_{r=r_c}=0,
\end{equation}
and yields, besides~\eqref{cf5} and~\eqref{cf6}, the third equation for the determination of the CPs\footnote{The global plus sings in the rhs's of~\eqref{cf3}, \eqref{cf4}, and~\eqref{cf8} would have been reversed had we chosen the signature ($+,\,-,\,-,\,-$) for~\eqref{0m1}.}
\begin{equation}\label{cf9}
    \al^2 = \frac{(\ln |g_{tt}|)_r\big|_{r=r_c}}{(\ln|g_{\ta\ta}|D)_r\big|_{r=r_c}}.
\end{equation}
This equation generalizes Eq.~(13) of Ref.~\cite{add} to the case of rotating solutions and rotating fluids.

Equations~\eqref{cf6} and~\eqref{cf9} cannot be handled further unless a stationary metric and an equation-of-state for the fluid are prescribed. This generally leads to a numerical analysis~\cite{aafj,aafs,add} to solve the non-linear system of equations~\eqref{cf6} and~\eqref{cf9} for the unknowns $r_c$ and $v_c$. We postpone this task to a subsequent work.

\section{Conclusion \label{secc}}
The accretion of non-degenerate and degenerate Fermi and Bose fluids have been modeled by the flow of a perfect fluid sharing with the stationary background metric the same symmetric properties. Conservation laws impose restrictions on the fluid flow at the end-points of the accessible region(s) to the fluid flow.

We have shown that the sort of ``critical exponents" ($x,\,y,\,z$) determine the behavior of the thermodynamic fields of the fluid as it approaches a horizon from within the region of the three-space where the Killing vector $\xi_t^{\mu}=(1,\,0,\,0,\,0)$ is timelike. This behavior is characterized by the presence of divergencies in the expressions of some fields along with the nullity of some other fields.

With the exception of the ultra-relativistic limit of the completely degenerate Fermi fluids, the presence of a power-law divergence in the expression of the pressure~\eqref{t13lib} seems not consistent with the other statistical models where the pressure either converges to a constant value or vanishes in the limit $r\to r_h$.

Critical flow in this model occurs for stationary black holes if initially the fluid rotates retrograde and for static black holes if initially the fluid is not rotating. At the critical point(s) the sound speed is larger than the radial three-velocity and smaller that the three-speed of the fluid element there.

The results apply to jets of particles too.

We have not assumed any theory of relativity, that is, we have just considered a general stationary metric with two Killing vectors. Hence the results apply to all black holes of all theories of gravitation, provided their metrics are endowed with the same properties of the general stationary metric we considered in this work.

%\section*{Acknowledgment} ?????

\section*{Appendix: Pressure gradient in the vicinity of the horizons\label{secaa}}
\renewcommand{\theequation}{A.\arabic{equation}}
\setcounter{equation}{0}
The purpose of this appendix section is to derive Eq~\eqref{t10li} giving the pressure gradient as the fluid approaches a horizon. First, we reverse Eqs.~\eqref{0m7} to express $u^{\mu}$~\eqref{m6} in terms of ($v,\,u$) and the metric components
\begin{multline}\label{ap1}
u^t=\gamma \sqrt{\frac{|g_{\phi\phi}|}{D}},\quad u^{r}=\frac{\gamma v}{\sqrt{|g_{rr}|}},\\
u^{\phi}=\frac{\gamma u}{\sqrt{|g_{\phi\phi}|}}+\frac{\gamma\sqrt{|g_{\phi\phi}|}}{\sqrt{D}}~\om,
\end{multline}
where we have omitted to write $u^{\ta}=\gamma\hat{u}/\sqrt{|g_{\ta\ta}|}$, which is identically zero in the $\ta=\pi/2$ plane. Using~\eqref{t8li} and~\eqref{t9li} we arrive at the following behaviors in the vicinity of the horizons
\begin{multline}\label{ap2}
\frac{\gamma}{\sqrt{D}}\simeq \frac{|r-r_h|^{-\frac{x+y}{2}}}{\sqrt{C_vC_D}},\quad \gamma u\simeq \frac{C_u|r-r_h|^{\frac{z-y}{2}}}{\sqrt{C_v}}.
\end{multline}
This shows that we can drop the first term in the expression of $u^{\phi}$. Using the fact that $D$ and $1/g_{rr}$ have the same behavior in the limit $r\to r_h$~\eqref{ta3} to yield $|g_{rr}|\simeq C_g|r-r_h|^{-x}$, we obtain
\begin{multline}\label{ap3}
\hspace{-3mm}u^{t}\simeq \frac{1}{\sqrt{C_vC_D}}\sqrt{|g_{\phi\phi}|}~|r-r_h|^{-\frac{x+y}{2}},\; u^r\simeq \frac{1}{\sqrt{C_vC_g}}~|r-r_h|^{\frac{x-y}{2}}\\
u^{\phi}\simeq  \frac{1}{\sqrt{C_vC_D}}\sqrt{|g_{\phi\phi}|}~\om |r-r_h|^{-\frac{x+y}{2}}.
\end{multline}
In Eqs.~\eqref{ap1} and~\eqref{ap2} we have assumed that $z>y$; if $z=y$ we just replace $C_v$ by $C_v-C_u^2$.

To obtain~\eqref{t10li} one may either evaluate $a^r=u^{r}\nabla_{r}u^{r}$ or
\begin{equation}
a^{\phi}=u^{r}\nabla_{r}u^{\phi}=u^{r}\partial_{r}u^{\phi}+\Gamma^{\phi}_{r\si}u^ru^{\si}.
\end{equation}
with $g^{\phi\phi}=-g_{tt}/D$, $g^{t\phi}=g_{t\phi}/D$, $u^{\phi}\simeq \om u^t$~\eqref{ap2}, and
\begin{multline}\label{ap4}
\Gamma^{\phi}_{r\si}u^{\si}=\tfrac{1}{2}(g^{\phi\phi}\partial_r g_{\phi\phi}+g^{t\phi}\partial_r g_{t\phi})u^{\phi}\\
+\tfrac{1}{2}(g^{\phi\phi}\partial_r g_{t\phi}+g^{t\phi}\partial_r g_{tt})u^{t},
\end{multline}
we obtain
\begin{align}
\label{ap5}&\Gamma^{\phi}_{r\si}u^ru^{\si}\simeq \frac{\partial_r(g_{t\phi}D)}{2Dg_{\phi\phi}}~u^ru^t,\\
\label{ap6}&u^{r}\partial_{r}u^{\phi}\simeq u^ru^t\partial_r\om +\om u^r\partial_ru^t.
\end{align}
To the leading order of approximation we drop the term proportional to $D\partial_rg_{t\phi}$ in~\eqref{ap5} since $D\partial_rg_{t\phi}/(g_{t\phi}\partial_rD)\sim |r-r_h|$ and the term proportional to $u^t$ in~\eqref{ap6} since $u^t/\partial_ru^t\sim |r-r_h|$. The law of motion~\eqref{lm} yields ($g^{\phi r}\equiv 0$)
\begin{equation}\label{ap7}
\nabla_rP\simeq \Big(\tfrac{g_{\phi\phi}}{2}~\partial_r\ln(D)-\partial_r\ln(u^t)\Big)nh.
\end{equation}
With $\partial_r\ln(D)\simeq x|r-r_h|^{-1}$~\eqref{t8li}, $\partial_r\ln(u^t)\simeq -(x+y)|r-r_h|^{-1}/2$~\eqref{ap3} [since $\partial_r\ln|g_{\phi\phi}|$ goes to a constant as $r\to r_h$], and $nh\sim |r-r_h|^{y-(x+z)/2}$~\eqref{t9li} we obtain
\begin{equation*}
\partial_{r}p\underset{(r\to r_h)}{\sim} |r-r_h|^{y-1-(x+z)/2},
\end{equation*}
which is Eq.~\eqref{t10li}.

Note that if the constraint $g_{\phi\phi}(r_h)x+x+y=0$ were imposed, the coefficient of $|r-r_h|^{-1}$ inside the parenthesis in~\eqref{ap7} would vanish and the gradient in the pressure and the pressure would behave as $$\partial_{r}p\underset{(r\to r_h)}{\sim} |r-r_h|^{y-(x+z)/2}$$ and $$p\underset{(r\to r_h)}{\simeq} p_h+\bar{C}_p|r-r_h|^{1+y-(x+z)/2}$$ where $\bar{C}_p$ is a constant, provided $y-(x+z)/2\neq -1$. The constraint $g_{\phi\phi}(r_h)x+x+y=0$ is, however, not generic, for it relates the parameters $x$ and $y$ to the metric value on the horizons. To keep the discussion as generic as possible we have dropped this case from our consideration.

%\section*{Appendix B: ??? $\pmb{???}$\label{secab}}
%\renewcommand{\theequation}{B.\arabic{equation}}
%\setcounter{equation}{0}

%\newpage


\begin{thebibliography}{99}
\bb{Dur}R.H.~Durisen,
%\emph{Accretion of rotating fluids by barytropes: Numerical results for white-dwarf models},
Astrophys. J. \textbf{213}, 145 (1977)

\bb{Matteo}T.~Di Matteo, S.W.~Allen, A.C.~Fabian, A.S.~Wilson, and A.J.~Young,
%\emph{Accretion onto the Supermassive Black Hole in M87},
Astrophys. J. \textbf{582}, 133 (2003)

\bb{Russell}H.R.~Russell, A.C.~Fabian, B.R.~McNamara, and A.E.~Broderick,
%\emph{Inside the Bondi radius of M87},
MNRAS \textbf{451}, 588 (2015)

\bb{acc00}H.~Bondi,
%\emph{On spherically symmetrical accretion},
MNRAS \textbf{112}, 195 (1952)
%Mon. Not. R. astr. Soc. \textbf{112}, 195 (1952)

\bb{acc0}F.C.~Michel,
%\emph{Accretion of matter by condensed objects},
Astrophys. Space Sci. \textbf{15}, 153 (1972)

\bb{acc1}K.S.~Thorne, R.A.~Flammang, and A.N.~Zytkow,
%\emph{Stationary spherical accretion into black holes. I - Equations of structure},
MNRAS \textbf{194}, 475 (1981)

\bb{acc2}R.A.~Flammang,
%\emph{Stationary spherical accretion into black holes – II. Theory of optically thick accretion},
MNRAS \textbf{199}, 833 (1982)

\bb{Bertozzi}A.L.~Bertozzi,
%\emph{Heteroclinic Orbits and Chaotic Dynamics in Planar Fluid Flows},
SIAM J. Math. Anal. \textbf{19}, 1271 (1988)
%http://epubs.siam.org/doi/abs/10.1137/0519093

\bb{Shapiro}S.L.~Shapiro,
%\emph{Accretion onto Black Holes: The Emergent Radiation Spectrum},
Astrophys. J. \textbf{180}, 531 (1973)

\bibitem{basic}A.~Treves, L.~Maraschi, and  M.~Abramowicz,
%\emph{Basic elements of the theory of accretion},
Astronomical Society of the Pacific \textbf{100}, 427 (1988)

\bb{Chat}I.~Chattopadhyay and D.~Ryu,
%\emph{Effects of Fluid Composition on Spherical Flows Around Black Holes},
Astrophys. J. \textbf{694}, 492 (2009)

\bibitem{string}A.~Ganguly, S.G.~Ghosh, and S.D.~Maharaj,
%\emph{Accretion onto a black hole in a string cloud background},
Phys. Rev. D \textbf{90}, 064037 (2014)

\bibitem{aafj}A.~K.~Ahmed, M.~Azreg-A\"{\i}nou, M.~Faizal, and M.~Jamil,
Eur. Phys. J. C \textbf{76}, 280 (2016)

\bibitem{aafs}A.~K.~Ahmed, M.~Azreg-A\"{\i}nou, S.~Bahamonde, S.~Capozziello, and M.~Jamil,
Eur. Phys. J. C \textbf{76}, 269 (2016)

\bibitem{add}M.~Azreg-A\"{\i}nou,
%\emph{Cyclic and heteroclinic flows near general static spherically symmetric black holes: Semi-cyclic flows -- Addendum and corrigendum},
Eur. Phys. J. C \textbf{77}, 36 (2017)

\bibitem{houches}V.~Beskin, G.~Henri, F.~Menard, G.~Pelletier, and J.~Dalibard (Eds.), \textit{Accretion discs, jets and high energy phenomena in astrophysics}, Les Houches Session LXXVIII, Series Volume 78 (Springer Berlin Heidelberg, Berlin 2003)
%http://link.springer.com/book/10.1007%2Fb80353

\bibitem{w2}M.~Azreg-A\"{\i}nou,
%\emph{Wormhole solutions sourced by fluids, II: three-fluid two-charged sources},
Eur. Phys. J. C \textbf{76}, 7 (2016)

\bibitem{Frank}J.~Frank, A.~King, and D.~Raine,
{\it Accretion Power in Astrophysics}, 3rd ed. (Cambridge University Press: NY 2002)

\bibitem{MHD}V.S.~Beskin,
{\it MHD Flows in Compact Astrophysical Objects}, (Springer-Verlag Berlin Heidelberg, Berlin 2010)

\bibitem{Rezzolla}L.~Rezzolla and O.~Zanotti,
{\it Relativistic Hydrodynamics}, (Oxford University Press: NY 2013)

\bibitem{Ellis}G.F.R.~Ellis, R.~Maartens, and M.A.H.~MacCallum, \textit{Relativistic Cosmology}, (Cambridge University Press, Cambridge 2012)

\bibitem{Kafka}P.~Kafka and P.~M\'{e}sz\'{a}ros,
%\emph{???},
Gen. Relativ. Grav. \textbf{7}, 841 (1976)

\bibitem{Gillman}A.W.~Gillman and R.F.~Stellingwerf,
%\emph{???},
Astrophys. J. \textbf{240}, 235 (1980)

\bibitem{GR}G.W.~Gibbons and D.A.~Rasheed,
%\emph{????},
Nucl. Phys. B \textbf{476}, 515 (1996)

\bibitem{phantom}G.~Cl\'ement, J.C.~Fabris and M.E.~Rodrigues,
%\emph{Phantom black holes in Einstein-Maxwell-dilaton theory},
Phys. Rev. \textbf{D 79}, 064021 (2009)

\bibitem{multi} M.~Azreg-A\"{\i}nou, G.~Cl\'ement, J.C.~Fabris, and M.E.~Rodrigues,
%\emph{Phantom black holes and sigma models},
Phys. Rev. {\bf D 83}, (2011) 124001

\bibitem{regular1}J.M.~Bardeen, ``Non-singular general relativistic gravitational collapse," in \textit{Proceedings of the 5th International Conference on Gravitation and the Theory of Relativity}, Edited by V. A.~Fock \textit{et al}. (Tbilisi University Press, Tbilisi, Georgia, 1968), p. 174
%\textsl{Non-singular general-relativistic gravitational collapse}, p. 174.
%J. Bardeen, "Non-singular general relativistic gravitational collapse," in Proceedings of the 5th International Conference on Gravitation and the Theory of Relativity, Ed. V. A. Fock et al., Tbilisi Unversty Press, Tbilisi, Georgia, September 1968.
%http://www.isgrg.org/pastconfs.php

\bibitem{reg}E.~Ay\'on-Beato and A.~Garc\'{\i}a,
%\emph{Regular black hole in general relativity coupled to nonlinear electrodynamics},
Phys. Rev. Lett. \textbf{80}, 5056 (1998)

\bibitem{gen2}M.~Azreg-A\"{\i}nou,
%\emph{Generating rotating regular black hole solutions without complexification},
Phys. Rev. D \textbf{90}, 064041 (2014)

\bibitem{Muller}I.~M\"{u}ller and T.~Ruggeri,
{\it Rational Extended Thermodynamics}, (Springer-Verlag: New York 1998)

\bibitem{Kremer}C.~Cercignani and G.M.~Kremer,
{\it The Relativistic Boltzmann Equation: Theory and Applications}, (Birkh\"{a}user: Boston 2002)

\bibitem{app1}C.~Cercignani,
{\it The Boltzmann equation and its applications}, (Springer: New York 1988)

\bibitem{app2}S.~Chandrasekhar,
{\it An introduction to the study of stellar structure}, (Dover: New York 1958)

\bibitem{Huang}K.~Huang,
{\it Statistical Mechanics}, (John Wiley: New York 1987)

\bibitem{hand}M.~Abramowitz and I.A.~Stegun (Eds.),
{\it Handbook of Mathematical Functions: with Formulas, Graphs, and Mathematical Tables}, (Dover: New York 1972)

\bibitem{1}M.~Ba\~{n}ados, J.~Silk, and S.M.~West,
%\emph{???},
Phys. Rev. Lett. \textbf{103}, 111102 (2009)

\bibitem{Gravitation}C.W.~Misner, K.S.~Thorne, and J.A.~Wheeler,
{\it Gravitation}, (W. H. Freeman: San Francisco 1973)

\end{thebibliography}
\end{document}